%% file: reflection.tex
\documentclass[acmtog, timestamp]{acmart}

\renewcommand\footnotetextcopyrightpermission[1]{} 
\def\BibTeX{{\rm B\kern-.05em{\sc i\kern-.025em b}\kern-.08emT\kern-.1667em\lower.7ex\hbox{E}\kern-.125emX}}

\setcopyright{none}

\setcitestyle{square}

\input{pre}

\begin{document}

\title{Least-Squares Affine Reflection Using Eigen Decomposition}

\author{Alec Jacobson}
\affiliation{
	\institution{University of Toronto}
	\streetaddress{40 St. George Street}
 	\city{Toronto}
 	\state{ON}
 	\postcode{M5S 2E4}
 	\country{Canada}}
\email{jacobson@cs.toronto.edu}



\begin{abstract}
  This note summarizes the steps to computing the best-fitting affine reflection
  that aligns two sets of corresponding points.\footnote{
I have attempted to \emph{maximize} similarity to the article by
\citet{SorkineRabinovichSVD} in order to highlight similarities in the
  mathematics.}
\end{abstract}

\maketitle

\section{Problem Statement}

\begin{wrapfigure}[10]{r}{0.80in}
  \includegraphics[trim=6.0mm 0mm 0mm 5mm,width=\linewidth]{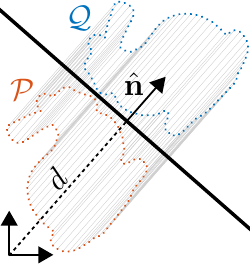}
\end{wrapfigure}
Let $\P = \{\p_1, \dots, \p_m\}$ and 
$\Q = \{\q_1, \dots, \q_m\}$ 
contain sets of $m$ corresponding points
$\p_i,\q_i∈\R^D$. We wish to find an affine reflection (i.e., reflection
across a hyperplane) that optimally aligns the two sets in the least squares
sense. That is, we seek the unit normal vector ($\n∈\R^D,‖\n‖=1$) and scalar
distance from the origin ($d∈\R$) describing a hyperplane ($\n⋅\p = d$) such
that 
\begin{align}
  (\n,d) = \argmin_{\n∈\R^d,‖\n‖=1,d∈\R} \ \sum_{i=1}^m \left‖ (\p_i-2(\p_i⋅\n - d)\n ) - \q_i \right‖².
\end{align}

\section{Computing the scalar distance to origin}
Fixing $\n$, denote $F(d) = 
\sum_{i=1}^m \left‖ (\p_i-2(\p_i⋅\n - d)\n ) - \q_i \right‖²
  $. We can find the optimal scalar term $d$ by taking the
derivative of $F$ with respect to $d$ and searching for its roots:

\begin{align}
  0 & = \frac{∂F}{∂d} =  \sum_{i=1}^m 4 \n^\top\left((\p_i-2(\p_i⋅\n - d)\n ) - \q_i \right), \\
    & = \sum_{i=1}^m 
    8 d \n^\top \n
    + 4 \n^\top \left(\p_i-2\n\n^\top\p_i  - \q_i \right)
  \intertext{\emph{(recalling that $\n^\top \n =‖\n‖ =  1$)}}
    & = 
    8 m d -  4 \n^\top \left(\sum_{i=1}^m \p_i + \q_i\right).
  \label{equ:dfdd}
\end{align} 
Let us introduce $\c ∈ \R^d$ to represent the centroid of \emph{all points}:
\begin{flalign}
    \text{\emph{Step 1}} && \c = \frac{1}{2m}\left(\sum_{i=1}^m  \p_i+
    \sum_{i=1}^m\q_i\right). &&
\end{flalign}
Substituting these into \refequ{dfdd}, we can express the optimal scalar $d$ in
terms of $\c$ and the (yet unknown) optimal $\n$:
\begin{flalign}
  \text{\emph{Step 5}} && 
  d = \c⋅\n. &&
\end{flalign}
In other words, the optimal scalar term ensures that the combined centroid of
the two sets lies on the reflective plane.  Or, equivalently, that the centroid
of $\P$ reflects to the centroid of $\Q$.

We can now substitute this optimal $d$ into our original objective function:
\begin{align}
  \sum_{i=1}^m \left‖ (\p_i-2(\p_i⋅\n - \c⋅\n)\n ) - \q_i \right‖².
\end{align}
Rearranging terms and injecting $\c-\c$ we can write this as
\begin{align}
  \sum_{i=1}^m \left‖ (\p_i-\c)-2((\p_i-\c)⋅\n) \n - (\q_i-\c) \right‖².
\end{align}

We can thus concentrate on computing the reflection plane normal $\n$ by
restating the problem such that the scalar term is zero (i.e., the plane passes
through the origin defining a linear reflection). Introduce the vectors from
each point to the combined centroid:
\begin{flalign}
  \text{\emph{Step 2}} && 
  \x_i = \p_i - \c \ \text{ and } \ \y_i = \q_i - \c. &&
\end{flalign}

So now we can look for the optimal unit normal such that:
\begin{align}
  \label{equ:nopt}
  \n = \argmin_{\n∈\R^D,‖\n‖=1} \ \sum_{i=1}^m \left‖ (\x_i-2(\x_i⋅\n) \n) - \y_i \right‖².
\end{align}

\section{Computing the unit normal vector}
Let us expand and simplify the term in the summation of \refequ{nopt}, removing
constants with respect to $\n$:
\begin{align}
  \left‖ (\x_i-2(\x_i⋅\n) \n) - \y_i \right‖² =  \\
   ‖\x_i‖² 
  -4\x_i^\top \n \n^\top(\x_i-\y_i)
  +4\x_i^\top \n \n^\top \x_i 
  + ‖\y_i‖² = \\
  \n^\top \x_i^\top \y_i \n \text{\emph{ (up to constants)}}.
\end{align}
Summing over these terms our optimization problem reduces to
\begin{align}
  \n = \argmin_{\n∈\R^D,‖\n‖=1} \ 
  \n^\top 
 \left(\sum_{i=1}^m \x_i \y_i^\top \right)
  \n.
\end{align}
By introducing
\begin{flalign}
  \text{\emph{Step 3}} && 
  \B = \sum_{i=1}^m \x_i \y_i^\top\ \text{ and } \ \A =  \frac{1}{2}(\B +
  \B^\top), &&
\end{flalign} we can further reduce this problem to 
\begin{align}
  \n = \argmin_{\n∈\R^D,‖\n‖=1} \n^\top \A \n.
\end{align}
This is the variational characterization of an eigen problem.
The optimal $\n$ is the eigenvector corresponding to the smallest eigenvalue:
\begin{flalign}
  \text{\emph{Step 4}} && \A \n = λ_\text{min} \n. &&
\end{flalign}
Revisiting our derivations we can identify the five \emph{Step}s necessary to
compute the best-fit affine reflection parameters $\n$ and $d$.

\bibliographystyle{ACM-Reference-Format}
\bibliography{references}

\end{document}

%% file: pre.tex
\usepackage{microtype} 
\usepackage{xspace}

\usepackage{breakcites}
\usepackage{ellipsis} 
\usepackage[mathletters]{ucs}
\usepackage[utf8x]{inputenc}
\usepackage{mathtools}
\usepackage{amsbsy}
\usepackage{upgreek}
\usepackage{dblfloatfix}

\usepackage{tabularx}
\usepackage{booktabs}

\usepackage{amsmath}

\usepackage{amsfonts}
\usepackage{amssymb}

\let\mat = \mathbf
\usepackage{xfrac}

\usepackage{amsmath}
\usepackage{amssymb}    
\usepackage{cancel}
\usepackage[T1]{fontenc}

\newcommand{\R}{\mathbb{R}}
\newcommand{\Q}{\mathcal{Q}}
\newcommand{\vc}[1]{\mathbf{#1}}

\renewcommand{\c}{\vc{c}}

\newcommand{\n}{\hat{\vc{n}}}

\newcommand{\p}{\vc{p}}
\newcommand{\q}{\vc{q}}

\newcommand{\x}{\vc{x}}

\newcommand{\y}{\vc{y}}

\newcommand{\A}{\mat{A}}
\newcommand{\B}{\mat{B}}

\renewcommand{\P}{\mathcal{P}}

\newcommand{\argmin}{\mathop{\text{argmin }}}

\DeclareUnicodeCharacter{8214}{\|}
\newcommand{\refequ}[1] {Equation~(\ref{equ:#1})}
\usepackage{wrapfig}

%% file: reflection.bbl

\begin{thebibliography}{1}


\ifx \showCODEN    \undefined \def \showCODEN     #1{\unskip}     \fi
\ifx \showDOI      \undefined \def \showDOI       #1{#1}\fi
\ifx \showISBNx    \undefined \def \showISBNx     #1{\unskip}     \fi
\ifx \showISBNxiii \undefined \def \showISBNxiii  #1{\unskip}     \fi
\ifx \showISSN     \undefined \def \showISSN      #1{\unskip}     \fi
\ifx \showLCCN     \undefined \def \showLCCN      #1{\unskip}     \fi
\ifx \shownote     \undefined \def \shownote      #1{#1}          \fi
\ifx \showarticletitle \undefined \def \showarticletitle #1{#1}   \fi
\ifx \showURL      \undefined \def \showURL       {\relax}        \fi
\providecommand\bibfield[2]{#2}
\providecommand\bibinfo[2]{#2}
\providecommand\natexlab[1]{#1}
\providecommand\showeprint[2][]{arXiv:#2}

\bibitem[\protect\citeauthoryear{Sorkine-Hornung and
  Rabinovich}{Sorkine-Hornung and Rabinovich}{2016}]%
        {SorkineRabinovichSVD}
\bibfield{author}{\bibinfo{person}{Olga Sorkine-Hornung} {and}
  \bibinfo{person}{Michael Rabinovich}.} \bibinfo{year}{2016}\natexlab{}.
\newblock \bibinfo{title}{Least-Squares Rigid Motion Using SVD}.
\newblock
\newblock
\newblock
\shownote{Technical note.}


\end{thebibliography}
